\documentstyle[epsf,sprocl]{article}
\def\figstki{
\begin{figure}[t] 
\begin{center}
\leavevmode
 \epsfxsize=0.49\hsize \epsfbox{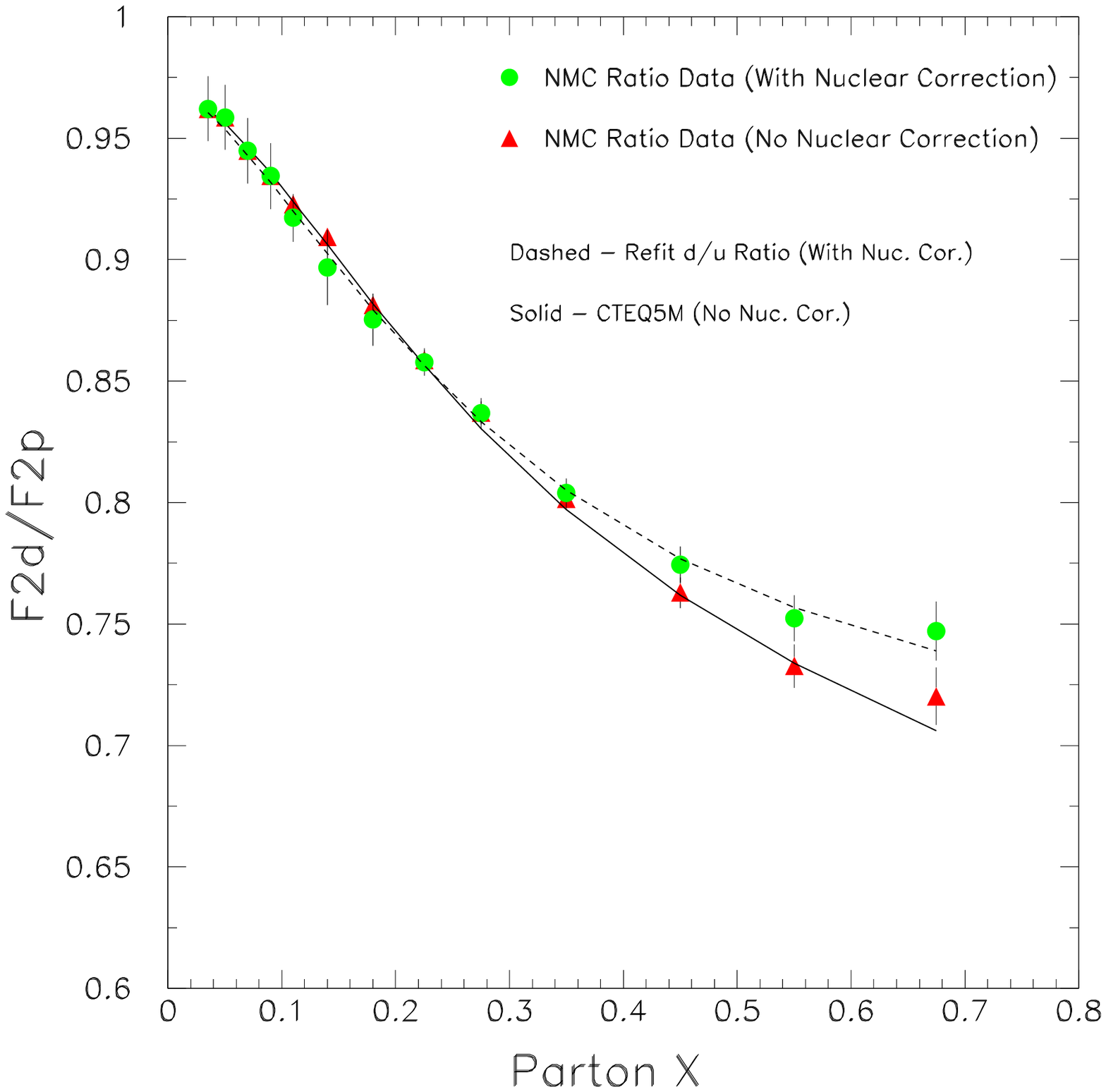} 
 \hfil
 \epsfxsize=0.49\hsize \epsfbox{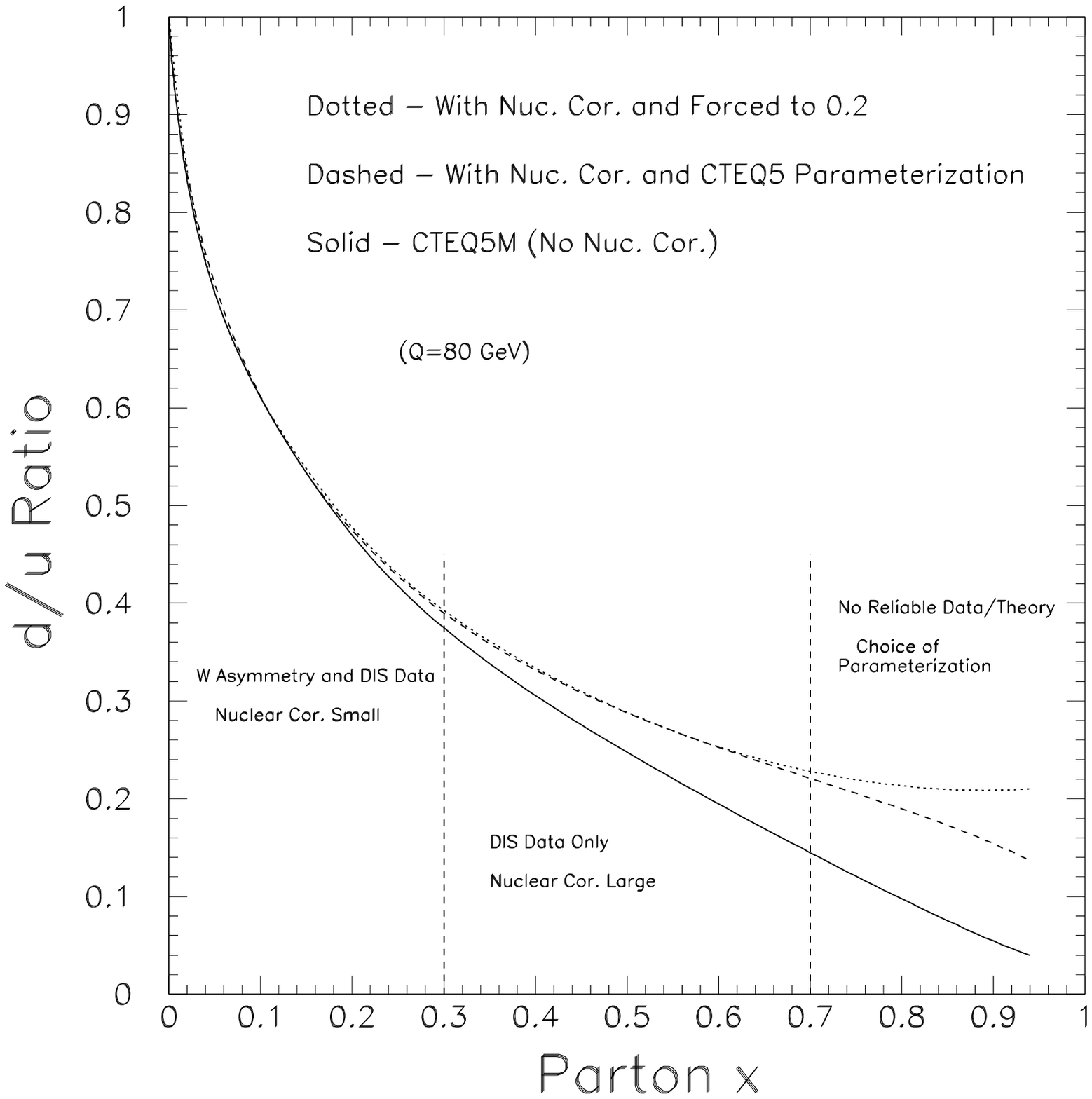} 
\vskip -10pt
\caption{
 a) The measured ratio of muon scattering off deuterium and hydrogen
targets, from the NMC experiment, is shown with and without the
nuclear corrections described in the text. The (lower) solid line is
CTEQ5M which was fit to the data with no nuclear corrections, while
the (upper) dashed line is a fit to the data with the nuclear
corrections, altering the $d/u$ ratio. 
 b) The $d/u$ ratio is shown for
the three global fits described in the text. Also shown are the three
different regions of $x$ and the relevant measurements in each region:
$i)$~DIS and $W$ Asymmetry data for $x < 0.3$, 
$ii)$~DIS only for $0.3 < x < 0.7$, 
and
$iii)$~no data for $0.7 < x$. 
\label{fig:stki} 
}
\vskip -20pt
\end{center}
\end{figure}
}
\def\figstkii{
\begin{figure}[t] 
\begin{center}
\leavevmode
 \epsfxsize=0.98\hsize \epsfbox{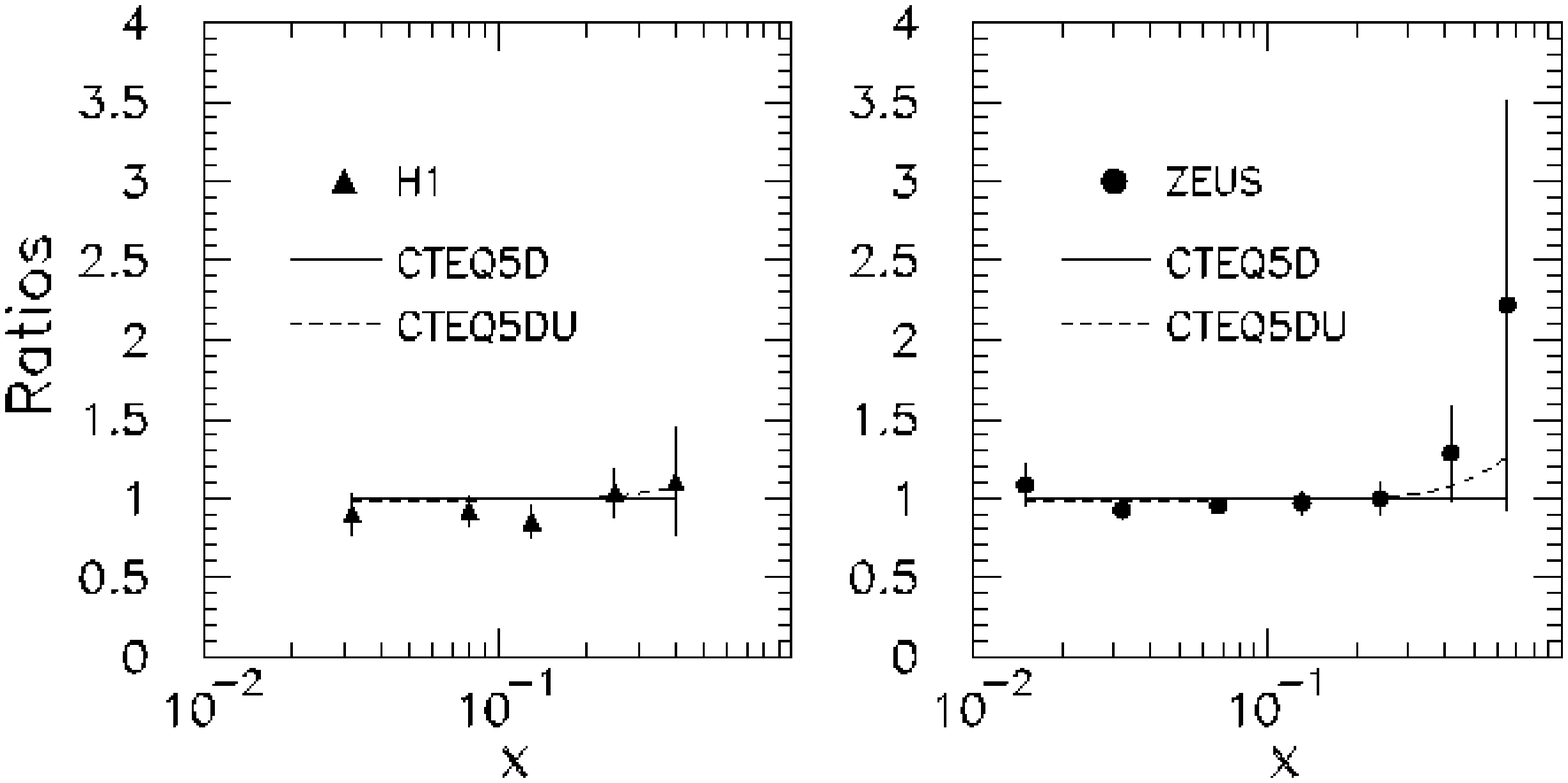} 
\vskip -10pt
\caption{
Positron-induced charged
current data from H1 and ZEUS, along with NLO QCD calculations using
parton distributions fit {\it with} (CTEQ5DU) and 
{\it without} (CTEQ5D) nuclear 
corrections to the fixed
target data (see text).
\label{fig:stkii} 
}
\vskip -20pt
\end{center}
\end{figure}
}
\def\gsim{\rlap{\lower 3.5 pt \hbox{$\mathchar \sim$}} \raise 1pt \hbox{$>$}}
\def\lsim{\rlap{\lower 3.5 pt \hbox{$\mathchar \sim$}} \raise 1pt \hbox{$<$}}

\def\fredcite#1{\,\cite{#1}}

\def\be{\begin{equation}}
\def\ee{\end{equation}}
\def\bea{\begin{eqnarray}}
\def\eea{\end{eqnarray}}

\begin{document}

\vskip 0.5in 

\begin{tabular}{l}
\null
\end{tabular}
\hfill
\begin{tabular}{r}
hep-ph/0007141  \\
SMU-HEP/00-11 \\
\end{tabular}

\vskip 1.5in 

\begin{center}
{\LARGE Parton Densitites at High--$x$}

\vskip 1.0in 

{\large
S.~Kuhlmann,$^a$ 
J.~Huston,$^b$ 
J.~Morfin,$^c$
F. Olness,$^d$
J.~Pumplin,$^b$ 
J.~F.~Owens,$^e$
W.~K.~Tung,$^b$ and 
J.~J.~Whitmore$^f$.
}

\vskip 0.25in 

$^a$Argonne National Laboratory, 
$^b$Michigan State University, 
$^c$Fermi National Accelerator Laboratory, 
$^d$Southern Methodist University
$^e$Florida State University, 
$^f$Pennsylvania State University, 

\vskip 0.5in

\begin{quote}
Reliable knowledge of parton distributions at large $x$ is crucial for
many searches for new physics signals in the next generation of
collider experiments.  Although these are generally well determined in
the small and medium $x$ range, it has been shown that their
uncertainty grows rapidly for $x > 0.1$. We examine the status of the
distributions in light of new questions that have been
raised about ``large-$x$'' parton distributions,
as well as recent measurements which have improved the parton
uncertainties. 

\end{quote}

\end{center}

\vfil

\noindent
{\small
Invited talk at 8th International Workshop on Deep Inelastic Scattering and QCD (DIS 2000),
Liverpool, England, 25-30 Apr 2000, 
presented by F.~Olness.
}


\title{PARTON DENSITIES AT HIGH--$x$\,\footnote{Presented by F.~Olness.}}

\author{
S.~KUHLMANN,$^a$ 
J.~HUSTON,$^b$ 
J.~MORFIN,$^c$
F. OLNESS,$^d$
J.~PUMPLIN,$^b$ 
J.~F.~OWENS,$^e$
W.~K.~TUNG,$^b$ and 
J.~J.~WHITMORE$^f$.
\\[5pt]
$^a$Argonne National Laboratory, 
$^b$Michigan State University, 
$^c$Fermilab,
$^d$Southern Methodist University
$^e$Florida State University, 
$^f$Pennsylvania State University, 
}



\maketitle

\abstracts{ \vskip 10pt 
Reliable knowledge of parton distributions at large $x$ is crucial for
many searches for new physics signals in the next generation of
collider experiments.  Although these are generally well determined in
the small and medium $x$ range, it has been shown that their
uncertainty grows rapidly for $x > 0.1$. We examine the status of the
distributions in light of new questions that have been
raised about ``large-$x$'' parton distributions,
as well as recent measurements which have improved the parton
uncertainties. 
}

\section{Introduction}

Four years ago the CDF collaboration reported\fredcite{cdfjet} an excess
of jet events at large transverse energy over perturbative Quantum
Chromodynamics (QCD) calculations.  A possible explanation for this
effect was a larger than expected gluon distribution\fredcite{cteqjet} at 
large $x$.  Three years ago the deep-inelastic scattering (DIS)
experiments at HERA reported a low statistics excess of events\fredcite{hera} at
large $Q^{2}$.  This led to speculation that part of this
excess could be attributed to a lack of knowledge of the quark
distributions\fredcite{highxq} at large $x$, and could possibly be
related to the jet events which are produced by a combination of quark
and gluon scattering.  Both excesses produced a large number of papers
about the possible implications for physics beyond the Standard Model,
emphasizing the need for much better knowledge of 
parton distributions\fredcite{scan}
at large $x$.

In the past few years there has been considerable progress towards
understanding some of the uncertainties in the individual measurements
that contribute to our knowledge of large-$x$ parton distributions
(PDFs); but, in some cases this has led to an \emph{increase} in the
uncertainty of the large-$x$ PDFs, rather than a reduction.  We will
review the recent analyses, and mention future measurements which
may help clarify this
situation.\footnote{For a 
comprehensive presentation of these issues, see Ref.~\fredcite{largex}.}

\figstki

\section{$d/u$ Ratio}

 The ratio of the density of down quarks to that of up quarks in the
proton has changed in the most recent CTEQ\fredcite{cteq5} 
and MRST\fredcite{mrst} analyses due to
the new $W$ lepton-asymmetry data from CDF,\fredcite{wasym} as well as the
NMC ratio measurement of deuterium/hydrogen scattering.\fredcite{nmc}
For many years the basic assumptions about the parameterization of
this ratio and the use of the DIS data have been relatively
unchallenged, but this has changed.  The two main reasons to question
these assumptions are: 1) the behavior of the $d/u$ ratio as
$x\rightarrow1$, and 2) possible nuclear binding effects in the
deuteron.  

To illustrate the different possibilities, a new series of fits were performed
within the context of the CTEQ5 global analysis.\fredcite{cteq5}   The nuclear
binding corrections were included as well as fits with a modified behavior of
$d/u$ as $x\rightarrow1$.  We find we can get a good fit to all the data with
neither correction, or with the nuclear binding corrections added but with any
$d/u$ behavior as $x\rightarrow1$.   
 Figure~\ref{fig:stki}a shows the NMC ratio data with and 
without the deuteron correction.
 The lower (solid) curve is CTEQ5M,  while the upper (dashed) curve is a new
fit to the corrected data, again with the standard CTEQ5 parameterization
which forces $d/u$ to zero as $x\rightarrow1$.   Both are good fits to the NMC
data, as  is a new third option (not shown since it lies precisely on the
dashed curve) which includes both the nuclear corrections and the changed $d/u$
parameterization.  
 Figure~\ref{fig:stki}b shows the $d/u$ ratio resulting from these three fits at $Q$=80 GeV
(there is very little evolution dependence in this ratio).  All three are
viable candidates for the $d/u$ ratio,  and the upper and lower ones could
quite reasonably be considered upper and lower bounds.  
Figure~\ref{fig:stki}b also
includes vertical lines to distinguish the three regions of $x$ involved in
this study,  and to help explain why the different effects can be treated as
independent.   For $x<0.3$ the W lepton-asymmetry data and the NMC ratio
data are both very precise \textit{and }the nuclear corrections to the NMC
data are insignificant.  The two measurements agree so the $d/u$ ratio is very
well constrained in this region.  Unfortunately the present W asymmetry data
end near $x=0.3$,   precisely where the nuclear corrections to the NMC data
become significant.  Therefore with any reasonably flexible parameterization
one can get a spread of $d/u$ ratios for $0.3<x<0.7$ (the middle region of the
plot) simply by changing the nuclear correction,  and still fitting the W
asymmetry and NMC data.   Finally for the largest $x$ values, we note that
the NMC data end near $x=0.7;$  therefore many different extrapolations to
$x\rightarrow1$ are possible, with or without nuclear corrections.  Clearly
the issues for the three different regions are quite independent.

\figstkii

It is worth noting that if $d/u\rightarrow0.2$ as $x\rightarrow1$,  then
there \textit{must }be some nuclear corrections in order to fit the NMC data.
 The previous discussion shows that the converse is not necessarily true.
  However if appreciable binding effects are present in the deuteron,  then
it is perhaps more natural for $d/u$ to go to a constant than to zero,  which
would require a fairly sharp downturn near $x=1$.   Assuming that $d/u$ does
not suddenly increase as $x\rightarrow1$,  this constant is unlikely to be
larger than 0.22,  since that is where the last NMC data point lies.   But
any constant between 0.05 and 0.2 would be a reasonable extrapolation and is
not constrained by present data.   

The best way to constrain the $d$ quark in the future is with high
luminosity HERA measurements of positron-induced charged current
interactions.    Figure~\ref{fig:stkii} show the most 
recent H1\fredcite{h1cc} and ZEUS\fredcite{zeuscc} charged current measurements. 
For the
H1 data the cuts are $Q^{2}>$ 1000 GeV$^{2}$and $y <$%
0.9, while for ZEUS the cut is $Q^{2}>200$ GeV$^{2}.$  They are compared to a
NLO QCD calculation using the standard CTEQ5D (DIS scheme) set of parton
distributions,  which are fit without the binding corrections to the NMC
data.   
This provides a good description of the data,
 although there is a hint of a low statistics excess in the ZEUS data.
   The dashed curves in the ratio plots are a second DIS scheme fit, which
we label CTEQ5DU, including binding corrections but with the CTEQ5
parameterization ($d/u \rightarrow0$) corresponding to the dashed curve (in
the $\overline{MS}$\ scheme) in Figure~\ref{fig:stki}b.   Since the data are below
$x<0.7$ the fits with $d/u\rightarrow0.2$ give the same result as CTEQ5DU in
this plot.   Parton distributions similar to the dashed and solid curves
were used to estimate the required luminosity to distinguish them.  The
result is that 500 pb$^{-1}$ of delivered positron luminosity (250 pb$^{-1}$
in each of the two experiments)\fredcite{zeus} is needed to achieve a 2 standard
deviation separation.   This is clearly a large data set but not impossible
with the forthcoming HERA upgrade.  We think it is vital that the HERA
program continue until this issue is settled.

We wish to thank M. Kuze, K. Nagano, and A. van Sighem for many useful
discussions and for the HERA luminosity estimate needed to determine the $d/u$ 
ratio.


\end{document}